\documentclass[%
 reprint,
 amsmath,amssymb,
 aps,
]{revtex4-1}

\usepackage{graphicx}
\usepackage{dcolumn}
\usepackage{bm}
\usepackage{wrapfig}
\usepackage{color}

\newcommand*{\bs}{\boldsymbol}

\begin{document}


\title{Mapping gravitational lensing of the CMB 
using local likelihoods}

\author{Ethan Anderes}
\thanks{Supported by NSF grant 1007480.}
\affiliation{%
Statistics Department\\ University of California, Davis, CA 95616}%

\author{Lloyd Knox}%
\thanks{Supported by NSF grant 0709498.}
 \affiliation{%
Physics Department\\ University of California, Davis, CA 95616
}%

\author{Alexander van Engelen}%
\thanks{Ph.~D.~candidate, McGill University.}
 \affiliation{%
Physics Department\\  McGill University, Montr\'eal H3A 2T8
}%


\begin{abstract}
We present a new estimation method for mapping the gravitational lensing potential from observed CMB intensity and polarization fields. Our method uses Bayesian techniques to estimate  the average  curvature of the potential over small local regions. These local curvatures are then used to construct an estimate of a low pass filter of the gravitational potential. By utilizing Bayesian/likelihood methods one can easily overcome problems with  missing and/or non-uniform pixels and problems with partial sky observations (E and B mode mixing, for example). Moreover, our methods are local in nature which allow us to easily model spatially varying beams and are highly parallelizable. We note that our estimates do not rely on the typical Taylor approximation which is used to construct estimates of the gravitational potential by Fourier coupling. We present our methodology with a flat sky simulation under nearly ideal experimental conditions with a noise level of $1$ $\mu K$-arcmin for the temperature field, $\sqrt{2}$ $\mu K$-arcmin for the polarization fields, with an instrumental beam full width at half maximum (FWHM) of $0.25$ arcmin.
\end{abstract}

\maketitle


\section{Introduction}
\label{into}

Over the past decade the cosmic microwave background (CMB) has emerged as a fundamental probe of cosmology and astrophysics. In addition to the primary fluctuations of the early Universe, the  CMB contains signatures of the gravitational bending of CMB photon trajectories due to matter, called gravitational lensing.  Mapping this gravitational lensing is important for a number of reasons including, but not limited to, understanding cosmic structure, constraining cosmological parameters \cite{Kaplin, Smth2006} and detecting gravity waves \cite{knox2002, Kesden, SelH}.  
 In this paper we present a local Bayesian estimate that can  accurately map the gravitational lens  in high resolution, low noise measurements of the CMB temperature and polarization fields.

 There is extensive literature on estimating the lensing of the CMB (classic references include \cite{ZaldSel1999, HuOka2002,HiraSel2003b}) and some recent observational detections \cite{Smith2007,Hira2008}. 
 The current estimators in the literature can be loosely characterized into two types. The first type was initiated in \cite{ZaldSel1999} (see also \cite{SelZlad1999,Guzik2000})  and utilizes quadratic  combinations of the  CMB and its gradient to infer lens structure.   The optimal quadratic combinations were  then  discovered by \cite{Hu2001b, HuOka2002, OkaHu2003} and are generally referred to as \lq the quadratic estimator'. This is arguably the most popular estimate of the gravitational potential and uses a first order Taylor approximation to establish mode coupling in the Fourier domain which can be estimated to recover the gravitational potential (real space analogs to these estimators can be found in \cite{buncher, carv}). The second type is an approximate global maximum likelihood estimate and was developed in \cite{HiraSel2003a, HiraSel2003b}.

 Our method, in contrast, locally approximates a quadratic form for the gravitational potential  and estimates the coefficients  locally using Bayesian methods. The locally estimated coefficients are then globally {\it stitched together} to construct an estimate of a  low pass filter of the gravitational potential. 
 The local analysis allows us to avoid using the typical first order Taylor expansion for the quadratic estimator and  avoids the likelihood approximations used in global estimates.
   Moreover, we are able to easily handle  missing pixels, problems with partial sky observations (E and B mode mixing, for example), and spatially varying or asymmetric beams. The motivation for developing this estimate stems, in part, from current speculation that likelihood methods will allow superior mapping of the lensing structure (compared to the quadratic estimator) under low noise levels, and that global likelihood methods can be prohibitively computational intensive---indeed intractable---without significant  approximation.


\begin{figure*}
\includegraphics[height=6.9cm]{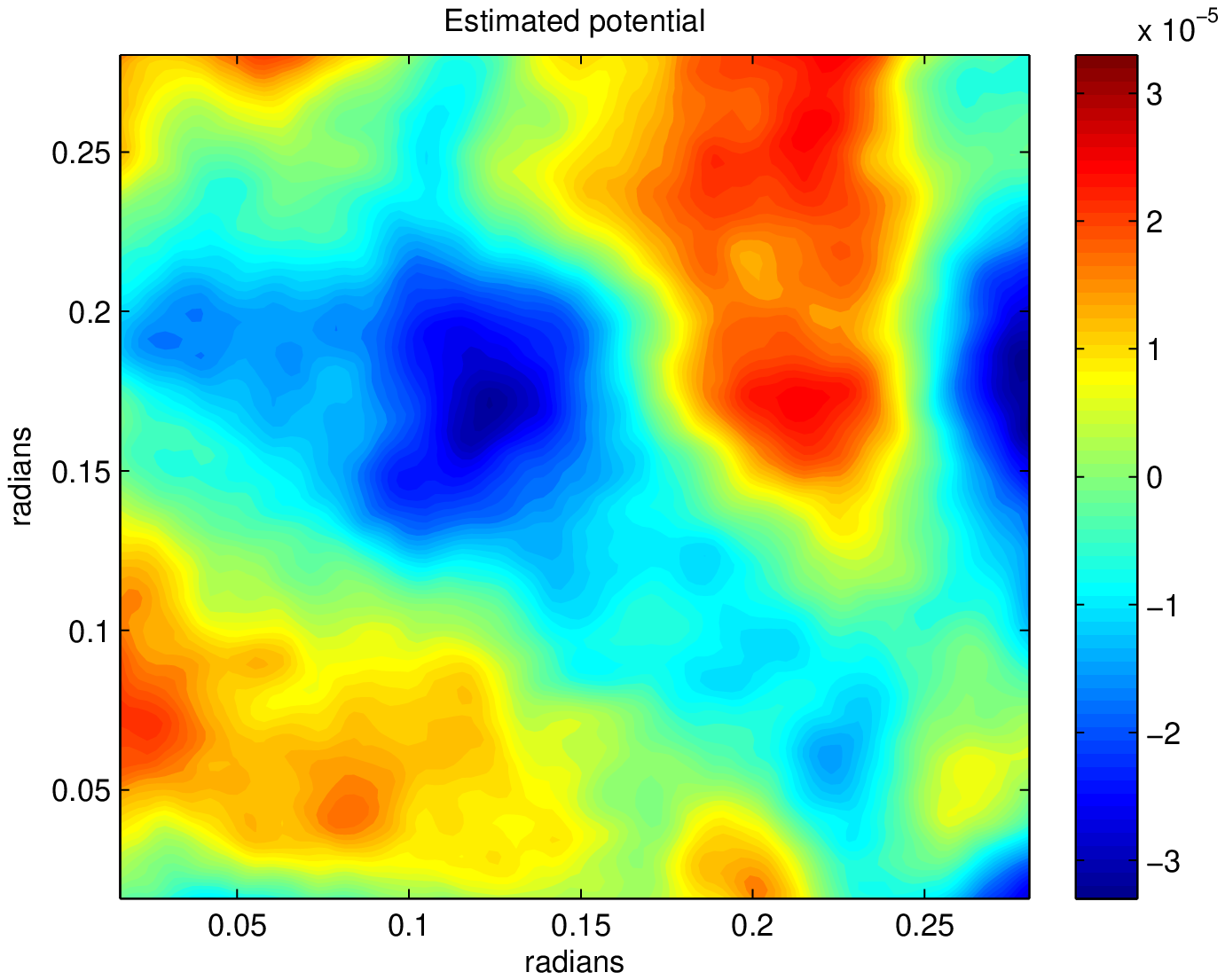}
\includegraphics[height=6.9cm]{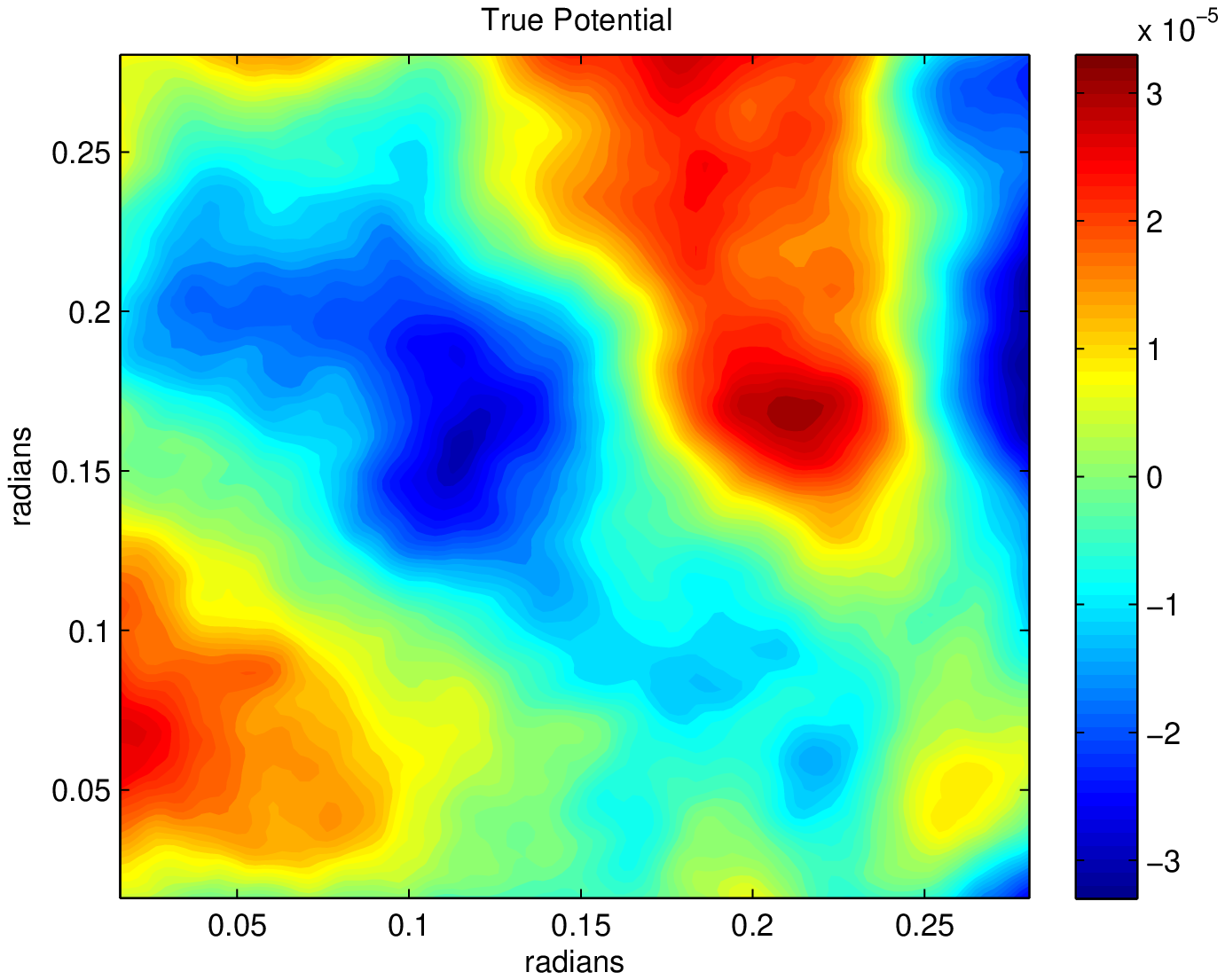}
\caption{{\it Left:} Estimated gravitational potential on a $17^o\times 17^o$ patch of the simulated flat sky.  {\it Right:} Input  gravitational potential used in the simulation. See Section \ref{into} and Appendix \ref{SimDets} for the simulation details.\label{fig1}}
\end{figure*}
 
\begin{figure}
\includegraphics[height=6.9cm]{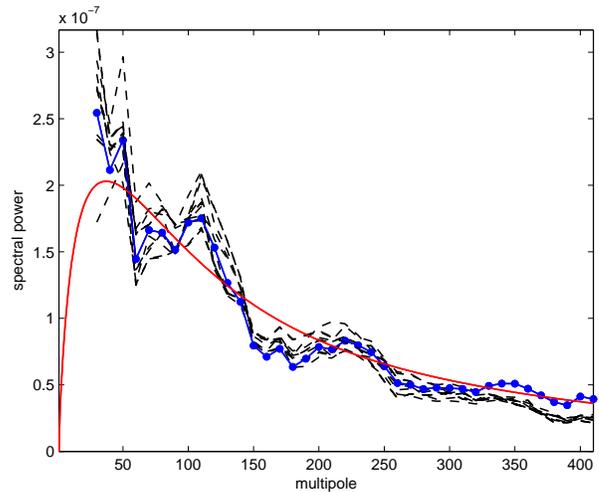}
\caption{Plot of the  projected mass power spectrum (red) along with  the estimated power spectrum using the true, but unknown, projected mass (blue). The dashed lines correspond to different estimates of the power spectrum using the estimated projected mass---with the local likelihood approach---for different CMB realizations but the same lensing potential realization. See Section \ref{into} and Appendix \ref{SimDets} for the simulation details. \label{Sect} }
\end{figure}

We illustrate our mapping methodology on a high resolution, low noise  simulation of the CMB temperature and polarization field on a  $17^\text{\,o}\times 17^\text{\,o}$ patch of the flat sky. This simulation is used throughout the paper to demonstrate findings and techniques. To get an overview of the performance of our method, Fig.~\ref{fig1} shows the estimated potential (left) from the simulated lensed CMB temperature and polarization field (observational noise levels are set at $1$ $\mu K$-arcmin for the temperature field, $\sqrt{2}$ $\mu K$-arcmin for the polarization fields,  with a beam FWHM of $0.25$ arcmin). The input gravitational potential is shown in the right diagram in Fig.~\ref{fig1}. 
The details of the simulation procedure can be found in Appendix \ref{SimDets}.   
It is clear from Fig.~\ref{fig1} that the mapping accurately traces the true, unknown gravitational potential. To get an idea of the noise of this reconstruction for different realizations of the CMB $+$ noise  we present Fig.~\ref{Sect} which shows 
 the different estimates of the projected matter power spectrum using the estimated projected mass---with the local likelihood approach---for 10 different CMB $+$ noise realizations (dashed lines) while keeping the gravitational potential in Fig.~\ref{fig1} fixed. The blue curve shows the estimated projected mass power spectrum if one had access to the true gravitational potential used in our simulations. Finally we plot the theoretical ensemble average projected mass power spectrum in red to get an idea of the magnitude of the errors in the mass reconstruction.  
  
\section{Local maximum {\it a posteriori} estimates of shear and convergence}
\label{FFF}

The CMB radiation in the flat sky limit can be expressed in term of the Stokes parameters $T,Q,U$ which measure total intensity $T(\bs x)$, and linear polarization $Q(\bs x)$ and $U(\bs x)$ with respect to some coordinate frame $\bs x=(x,y)\in\Bbb R^2$.
Instead of directly observing $T,Q,U$ we observe a remapping of the CMB due to the gravitational effect of intervening matter. This lensed CMB can be written $T(\bs x+\nabla \phi(\bs x)),Q(\bs x+\nabla \phi(\bs x))$ and $U(\bs x+\nabla \phi(\bs x))$ where  $\phi$ denotes the gravitational potential (see \cite{dod}, for example).

To describe our estimate of the gravitational potential, $\phi$,  first consider a small circular observation patch with diameter $\delta$ in the flat sky centered at some point $\bs x_0$, denoted $\mathcal N_\delta(\bs x_0)\subset \Bbb R^2$.
 Over this small region  we decompose $\phi$ into an overall local quadratic fit and error term 
\[\phi=q^\phi+\epsilon\] 
where $q^\phi$ is a local quadratic approximation of the potential $\phi$ with error term $\epsilon\equiv \phi-q^\phi$.  In what follows we estimate $q^\phi$, denoted $\hat q^\phi$, and associate this estimate with the neighborhood midpoint $\bs x_0$. Then we repeat this procedure for other local midpoints $\bs x_0$ throughout the observation window. After a shrinkage adjustment is made to the local estimates we show, in Section \ref{Rec},  how to {\it stitch} the estimates together to produce the final estimated potential $\hat\phi$ shown in Fig.~\ref{fig1}. 

Notice that as $\delta\rightarrow 0$ the expected magnitude of the error $\epsilon$ approaches zero. This  has the effect of improving the following Taylor approximation 
\begin{align}
T(\bs x+\nabla \phi(\bs x))&=T(\tilde{\bs  x}) + \nabla\epsilon(\bs x) \cdot \nabla T(\tilde{\bs  x}) \label{number2} +\cdots
\end{align}
for $\bs x\in\mathcal N_\delta(\bs x_0)$,
where we use the notation  $\tilde{\bs  x}\equiv {\bs  x}+\nabla q^\phi({\bs  x})$ (with a similar Taylor expansion for both $Q(\bs x +\nabla \phi(\bs x))$ and $U(\bs x +\nabla \phi(\bs x))$). Notice that  $\tilde{\bs  x}$  depends not only on ${\bs  x}$ but also the unknown coefficients of the quadratic term $q^\phi$.   We  briefly mention that these are related to the convergence $\kappa$ and shear $\gamma=\gamma_1+i \gamma_2$ of the gravitational lens by 
  \begin{align*} 
 \kappa  &\approx -(q^\phi_{xx} + q^\phi_{yy})/2 \\
\gamma_1 &\approx -(q^\phi_{xx} - q^\phi_{yy})/2\\
 \gamma_2&\approx - q^\phi_{xy} \end{align*}
 using the shear notation given in \cite{ZaldSel1999}.
Now when $\delta$ is sufficiently small we can truncate the expansion in (\ref{number2}) to get
\begin{equation}
\label{approx1}
\left[
 \begin{array}{c}
T({\bs  x}+\nabla \phi({\bs  x}))\\
  Q({\bs  x}+\nabla \phi({\bs  x})) \\
   U({\bs  x}+\nabla \phi({\bs  x})) 
  \end{array}
  \right]
  \approx
   \left[\begin{array}{c}
T(\bs x+\nabla q^\phi(\bs x))\\
  Q(\bs x+\nabla q^\phi(\bs x)) \\
   U(\bs x+\nabla q^\phi(\bs x)) 
 \end{array}
 \right]
 \end{equation}
 on the local neighborhood $\mathcal N_\delta(\bs x_0)$.
 By regarding  $q^\phi$ as unknown we can use the right hand side of (\ref{approx1})  to develop a likelihood for estimating  the coefficients of $q^\phi$.  Nominally $q^\phi$ has $6$ unknown coefficients for which to estimate. However,  we can ignore the linear terms in $q^\phi$ since the CMB temperature and the polarization are statistically invariant under the resulting translation in $\nabla q^\phi$.  Therefore, one can write  $q^\phi$ as $c_1(x-x_0)^2/2 + c_2(x-x_0)(y-y_0) +  c_3 (y-y_0)^2/2$ for unknown coefficients $c_1=q^\phi_{xx}, c_2=q^\phi_{xy},  c_3=q^\phi_{yy}$.
 
 An important probe of gravitational lensing from the CMB polarization is the creation of a curl-like B mode from the  lensing \citep{Kami,SelZlad1998}. We remark that a local quadratic approximation  in (\ref{approx1}) still has the power to detect this B mode power so that the local procedure is not blind to this information source.  To see this notice that a quadratic lensing potential remaps the coordinates by
\[ \tilde{\bs x}=\bs x + \left[\begin{array}{cc} q_{xx}^\phi &  q_{xy}^\phi \\ q_{xy}^\phi& q_{yy}^\phi\end{array}\right] ( \bs x -\bs x_0).\]
If we assume the original polarization $(Q(\bs x),U(\bs x))$ is curl free then the lensed polarization has curl given by
\begin{align*}
\text{curl}(Q(\tilde{\bs x}),U(\tilde{\bs x}))
&= -2\gamma_2U_x(\tilde{\bs x})  +\gamma_1 \bigl[Q_x(\tilde{\bs x})-U_y(\tilde{\bs x}) \bigr]. 
\end{align*}
Therefore the shear  parameter $\gamma$, and not the convergence $\kappa$, is what creates local  B-mode power. The dominant source of information for B-mode power is in the cross correlation between the lensed Stokes parameters $Q(\tilde{\bs x})$ and $U(\tilde{\bs x})$. This agrees with \cite{HuOka2002} that the E-B cross estimator provides optimal signal to noise under nearly ideal experimental conditions.

We finish this section with a remark on the accuracy of the Taylor approximation (\ref{number2}).
As the the signal to noise ratio increases and the pixel resolution improves one can shrink the local neighborhood $\mathcal N_\delta(\bs x_0)$ so the term $\epsilon$ becomes smaller (which improves the Taylor approximation).  However, as $\delta\rightarrow 0$, the fields $T,Q$ and $U$ become nearly linear and one may expect some loss of information from the shrinking power in $T,Q$ and $U$ at frequencies with wavelengths smaller than the neighborhood $\mathcal N_\delta(\bs x_0)$. It therefore may be statistically advantageous to artificially increase the neighborhood size while simultaneously increasing the order of the local polynomial fit $q^\phi$. Then, instead of recording the full polynomial fit at each midpoint $\bs x_0$, one can retain the second order derivatives  $q^\phi_{xx}(\bs x_0),q^\phi_{xy}(\bs x_0), q^\phi_{y}(\bs x_0)$ for estimates of $\kappa$ and $\gamma$. It is yet to be seen, however, what  $\delta$ and what polynomial order will be optimal for a given noise and resolution level. In Section \ref{Infor} we present an information metric for choosing the neighborhood size $\delta$ for the simulation specifics and for a quadratic polynomial $q^\phi$.

\subsection{The local posterior}
\label{lllike}

Using the Gaussian approximation of the CMB  along with the quadratic potential approximation given by (\ref{approx1}) we describe how to construct the likelihood as a function of the unknown quadratic coefficients in $q^\phi$. Let $\bs x_1,\ldots, \bs x_n$ denote the observation locations of the CMB within the local neighborhood $\mathcal N_\delta (\bs x_0)$ centered at $\bs x_0$. 
Using approximation (\ref{approx1}), the CMB observables in this local neighborhood are well modeled by white noise corruption of a convolved (by the beam) lensed intensity and polarization field. Let $\bs t, \bs q, \bs u$ denote $n$-vectors of observed CMB values at the corresponding pixel locations in $\mathcal N_\delta(\bs x_0)$ for the intensity $T$ and Stokes parameters $Q, U$, respectively. 
Using Gaussianity of the full vector of CMB observables, $\bs z=(\bs t^\dagger,\bs q^\dagger,\bs u^\dagger)^\dagger$, the log likelihood (up to a constant) as a function of the quadratic fit $q^\phi$ can be written
\begin{equation}
\label{MMTT}
\mathcal L(q^\phi|\bs z) = -\frac{1}{2}\bs z^\dagger \left(\Sigma_{q^\phi} +N \right)^{-1}\bs z -\frac{1}{2}\ln \det \left(\Sigma_{q^\phi}+ N \right)  
\end{equation}
where $\Sigma_{q^\phi}+N$ is the covariance matrix of the observation vector $\bs z$ (we use the subscript to emphasize the dependence on the unknown quadratic $q^\phi$), $N=\text{diag}\bigl (\sigma^2_T I,\sigma^2_Q I,\sigma^2_U I\bigr )$ is the noise covariance structure and $I$ is the $n\times n$ identity matrix. Notice that the noise structure does not depend on the unknown quadratic $q^\phi$. In the next section we will derive the exact form of the prior distribution on $q^\phi$, denoted $\pi(q^\phi)$, but briefly mention that the posterior distribution on $q^\phi$, which we maximize to estimate $q^\phi$, is
\begin{equation}
\label{FormPost}
p(q^\phi | \bs z)\propto e^{\mathcal L(q^\phi|\bs z)}\pi(q^\phi).  
\end{equation}

The entries of $\Sigma_{q^\phi}+N$ contain the covariances $\left\langle  t_k t_j\right\rangle_\text{CMB}$, $\left\langle  q_k q_j\right\rangle_\text{CMB}$, $\left\langle  u_k u_j\right\rangle_\text{CMB}$ and all cross covariances among $\bs t,\bs q,\bs u$ (we use $t_k$ to denote the $k^\text{th}$ entry of $\bs t$, for example).
Let $\varphi$ denote the instrumental beam  and $\sigma_T,\sigma_Q,\sigma_U$ denote the noise standard deviations of  $T,Q,U$ so that, for example, the $k^\text{th}$ entry of $\bs t$ is modeled as
\begin{equation}
\label{obbbsss} 
t_k\equiv  \int_{\Bbb R^2}\! d^2\bs x\,  \varphi(\bs x) T(\tilde{\bs x}_k-\tilde{\bs x}) + \sigma_T n_{k} 
\end{equation}
 where the  $n_k$'s are independent standard Gaussian random variables, $\tilde{\bs x}_k=\bs x_k +\nabla q^\phi(\bs x_k)$ and $\tilde{ \bs x}=\bs x +\nabla q^\phi(\bs x)$.  Note that this is an approximate model for $t_k$ based on (\ref{approx1}). In actuality, the $k^\text{th}$ temperature measurement is  $\int_{\Bbb R^2}\! d^2\bs x\,  \varphi(\bs x) T({\bs x}_k-{\bs x}+\nabla\phi({\bs x}_k-{\bs x})) + \sigma_T n_{k} $, but the linearity of $\nabla q^\phi$ allows us to write ${\bs x}_k-{\bs x}+\nabla\phi({\bs x}_k-{\bs x})\approx\text{constant}+ \tilde{\bs x}_k-\tilde{\bs x} $ on the small neighborhood $\mathcal N_\delta(\bs x_0)$.
 Under the assumption of zero $B$ mode, the spectral densities associated with  $Q, U$ can be written
\begin{align}
C^{Q}_{\bs \ell} &=C^{E}_{\bs \ell}  \cos^2(2\varphi_{\bs \ell}) \\
C^{U}_{\bs \ell} &= C^{E}_{\bs \ell} \sin^2(2\varphi_{\bs \ell}) \\
C^{QU}_{\bs \ell} &=  C^{E}_{\bs \ell} \cos(2\varphi_{\bs \ell}) \sin(2\varphi_{\bs \ell})
\end{align}
where $\tan(\varphi_{\bs \ell})=\ell_2/\ell_1$ and ${\bs \ell}=(\ell_1,\ell_2)\in\Bbb R^2$. 
Since one can write $\bs x+ \nabla q^\phi(\bs x)=M\bs x$  where the $M$ is a $2\time 2$ real matrix,
the sheared Stokes parameters $T(\tilde{\bs x}), Q(\tilde{\bs x})$ and $U(\tilde{\bs x})$ are stationary random fields with spectral densities given by $C^T_{M^{-1}\bs \ell}\det M^{-1}, C^Q_{M^{-1}\bs \ell}\det M^{-1}$ and $C^U_{M^{-1}\bs \ell}\det M^{-1}$, respectively. After adjusting for the beam (which is applied after lensing) the covariance between the observations in $\bs t$ can be written
\begin{align}
\left\langle  t_k t_j\right\rangle_\text{CMB}=\sigma^2_T \delta_{ij} + \int_{\Bbb R^2} \frac{d^2\bs \ell}{(2\pi)^2} e^{i\bs \ell\cdot (\bs x_k-\bs x_j)}  |\varphi(\bs \ell)|^2 \frac{C^T_{M^{-1}\bs \ell}}{\det M}.
\end{align}
The computations are similar to complete the entries of covariance matrix $\Sigma_{q^\phi}+N$.
At face value the above integral seems too computationally intensive  for every pair $\bs x_k-\bs x_j$. Moreover, to apply Newton type algorithms for maximizing the posterior (\ref{FormPost}) one needs to compute the derivatives of $\left\langle  t_k t_j\right\rangle_\text{CMB}$ with respect to elements of $M$. In Appendix \ref{Newt} we show that some of these computational challenges can be overcome by utilizing a single FFT to quickly compute the above integral for sufficient resolution in the argument $\bs x_k-\bs x_j$ to recover $\left\langle t_k t_j\right\rangle_\text{CMB}$ for all pairs $k,j$.

\subsection{Taylor truncation bias}
\label{truc}

The quadratic function  $q^\phi$ is defined  as the best least square fit of  $\phi$ over the neighborhood $\mathcal N_\delta (\bs x_0)$. The residual $\epsilon=\phi-q^\phi$, defined over $\mathcal N_\delta(\bs x_0)$, is nonstationary and will therefore not have a spectral density that diagonalizes the covariance structure. However, stationarity is a good approximation for order of magnitude calculations on the truncation error in (\ref{number2}). We  approximate the  spectral density of $\epsilon$ as an attenuated version of $C_\ell^\phi$ by arguing that  the quadratic fit  effectively removes the spectral power at wavelengths greater than $2\delta$.  
Reasoning similarly we expect the quadratic fit to have negligible impact on the spectral power at wavelengths smaller than $\delta$. By assuming the spectral power grows linearly in the intermediary spectral range, from zero at $\ell=\pi/\delta$ to $C^\phi_{2\pi/\delta}$ at $\ell=2\pi/\delta $, we obtain an approximate model for the spectral density of $\epsilon$
\[
C^\epsilon_{\bs \ell}\approx \min\Bigl\{1, \Bigl[\frac{\delta}{\pi} |\bs \ell|  - 1 \Bigr]^+\Bigr\}^2 {C_\ell^\phi}
\]
where $x^+$ denotes the positive part of the real number $x$.
Notice that the attenuation happens on the realizations of $\phi$, hence requiring the square on the low pass filter in the spectral density.
This implies that the second term in the Taylor expansion (\ref{number2}) has approximate spectral density
\[ C^{\nabla T(\tilde{\bs x}) \cdot\nabla\epsilon(\bs x) }_{\bs \ell} \approx \int \frac{d^2\boldsymbol \ell^\prime}{(2\pi)^2} (M^{-1}\boldsymbol \ell^\prime \cdot (\boldsymbol \ell-\boldsymbol \ell^\prime))^2 C^\epsilon_{\boldsymbol \ell-\boldsymbol \ell^\prime} \frac{C^{T}_{M^{-1}\boldsymbol\ell^\prime}}{\det M}. \]

\begin{figure}
\includegraphics[width=9cm]{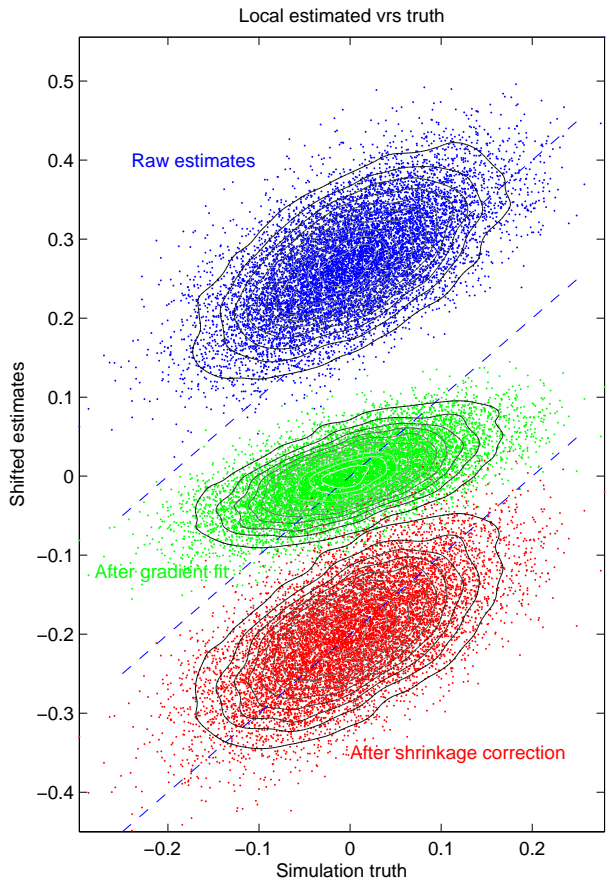}
\caption{Estimated values of $\nabla^2 \phi(\bs x_0)$, for each local neighborhood midpoint $\bs x_0$, plotted against  the simulation truth at different stages of the algorithm.   The blue points correspond to the raw estimates at each local neighborhood; The green points to the estimates after fitting a gravitational potential; The red points after a shrinkage correction. The $y$ coordinates of the blue points are shifted up by 0.2 and the red points are shifted down by 0.2 to fit on the same diagram. See Sections \ref{truc} and \ref{Rec} for  discussion.\label{fig2}}
\end{figure}

In our simulation we use a neighborhood diameter of $\delta=0.006$ radians (20.6 acrmin). This  diameter was chosen using the information criterion developed in Section \ref{Infor}. 
The corresponding approximate rms of $\nabla T(\tilde{\bs x}) \cdot\nabla\epsilon(\bs x)$ is $\sim 2.3\, \mu K$ with an order of magnitude reduction for the polarization field. Brute force simulation of $\left\langle \underset{\bs x_k\in\mathcal N_\delta(\bs x_0)}{\text{Mean}}\left\{T(\bs x_k+\nabla \phi(\bs x_k))-T(\tilde{\bs x}_k)\right\}^2\right\rangle_{CMB}^{1/2}$ yields a value closer to $\sim 3.6\,\mu K$,  
suggesting a reasonable  stationary approximation to $\epsilon$. 
These approximations show that the polarization truncation error is smaller (by an order of magnitude) than the simulation noise level $\sqrt{2}\,\mu K$-arcmin. However, the temperature truncation error is greater than the temperature noise level $1\,\mu K$.   A consequence is that the likelihood explains the additional high frequency  power in the observations (from the error term)  by adjusting the estimate of  $q^\phi$ to artificially magnify the convergence $\kappa$ estimates. Indeed, this bias seems  relatively constant and can be clearly seen in Fig.~\ref{fig2} in the top blue points. Each blue point corresponds to a local neighborhood: the $x$-coordinate representing the true  $\nabla^2 q^\phi$ associated with that neighborhood; the $y$-coordinate representing the estimated local value shifted up by $0.2$, i.e.\! $ \nabla^2\hat q^\phi+0.2$.
The bias of nearly $\sim 0.1$ above the top dashed blue line $y=x+0.2$, shows the effect of the additional high frequency power of the error term $\nabla T(\tilde{\bs x}) \cdot\nabla\epsilon(\bs x)$.  To adjust this, we subtract the overall mean of the local estimates, reasoning that the observation window is large enough at $17^\text{\,o}\times 17^\text{\,o}$ so that the overall mean of the true values $ q^\phi_{xx}, q^\phi_{xy}, q^{\phi}_{yy}$ is close to zero.
 For smaller observation windows it may be possible to estimate an overall quadratic fit to correct for this bias but we do not investigate that here.

\subsection{The prior $\pi(q^\phi)$}
\label{pprior}
The stationary approximation for $\epsilon$ also yields an approximation for the the prior distribution  of the local quadratic fit $q^\phi$ using the identity $q^\phi=\phi-\epsilon$. Since $\epsilon$ is well modeled by a high pass filter of $\phi$, the quadratic function $q^\phi$ can be modeled by the corresponding low pass filter
\[ q^\phi(\bs x)  \approx \int \frac{d^2\bs \ell}{2\pi} e^{i \bs x\cdot \bs \ell} \phi^\text{lp}(\bs \ell)\]
over $\bs x\in \mathcal N_{\delta}(\bs x_0)$, where 
$ \phi^\text{lp}(\bs \ell)\equiv   \min\Bigl\{1, \Bigl[2-\frac{\delta}{\pi} |\bs \ell|  \Bigr]^+\Bigr\} \phi(\bs \ell) $ which has spectral density $\min\Bigl\{1, \Bigl[2-\frac{\delta}{\pi} |\bs \ell|  \Bigr]^+\Bigr\}^2 {C_\ell^\phi} $. Therefore a natural candidate for the prior on the coefficients of $q^\phi$ are the random variables $\frac{\partial^2\phi^\text{lp}(0)}{\partial x_k\partial x_j}$ which are mean zero and Gaussian  with variances obtained by  the corresponding spectral moments of $\phi^\text{lp}$.  This prior is used on each local neighborhood $\mathcal N_{\delta}(\bs x_0)$ to derive the local maximum {\it a posteriori} estimate. For the simulation used in this paper, the neighborhood width was set to $\delta=0.006$ radians (20.6 arcmin) which gives prior variances $0.0023, 0.0008, 0.0023$ for  $q_{xx}^\phi, q_{xy}^\phi$ and $q_{yy}^\phi$, respectively (the only nonzero cross covariance is between $q_{xx}^\phi$ and $q_{yy}^\phi$ and is $0.0008$).


\begin{figure*}
\includegraphics[height=6.7cm]{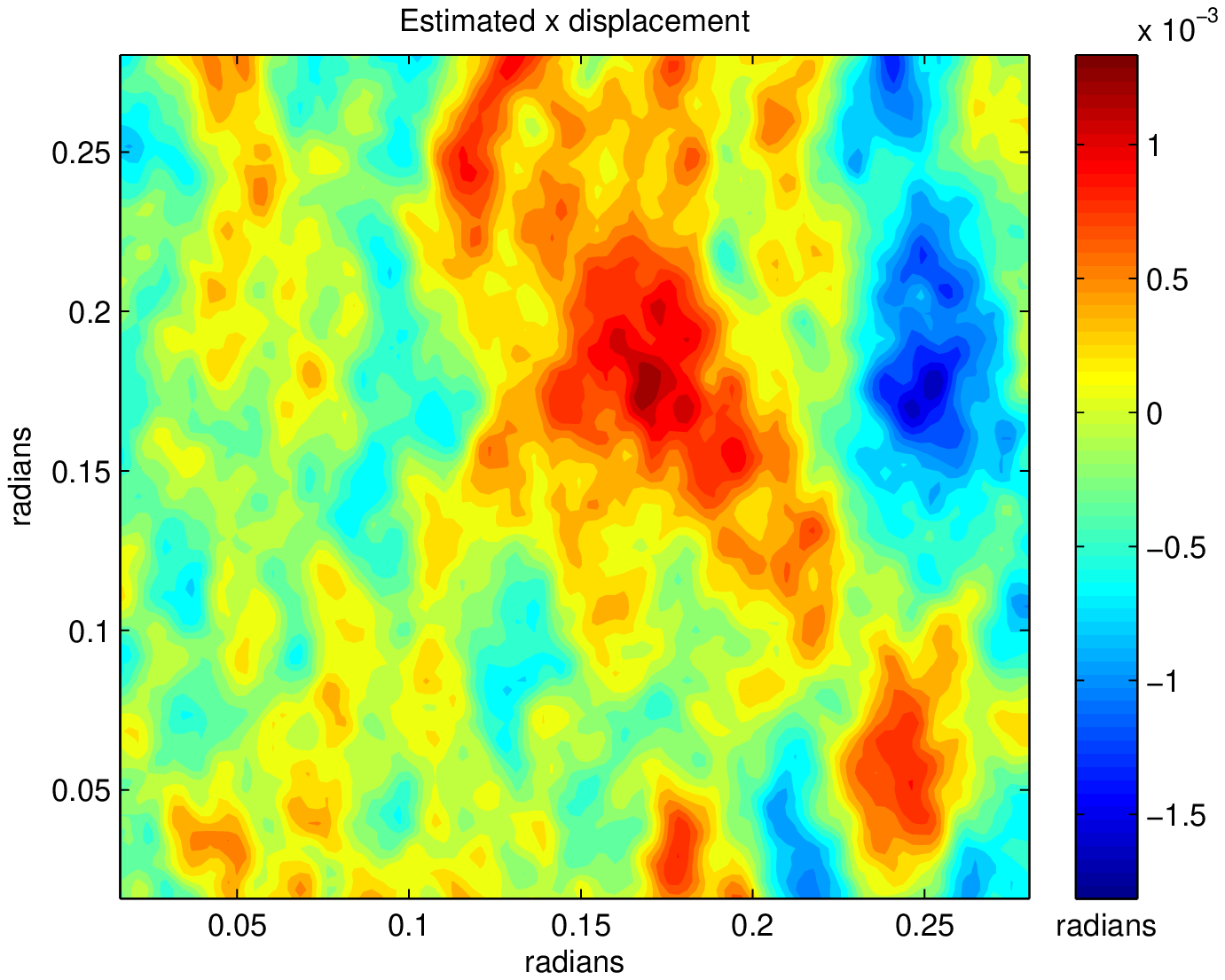}
\includegraphics[height=6.7cm]{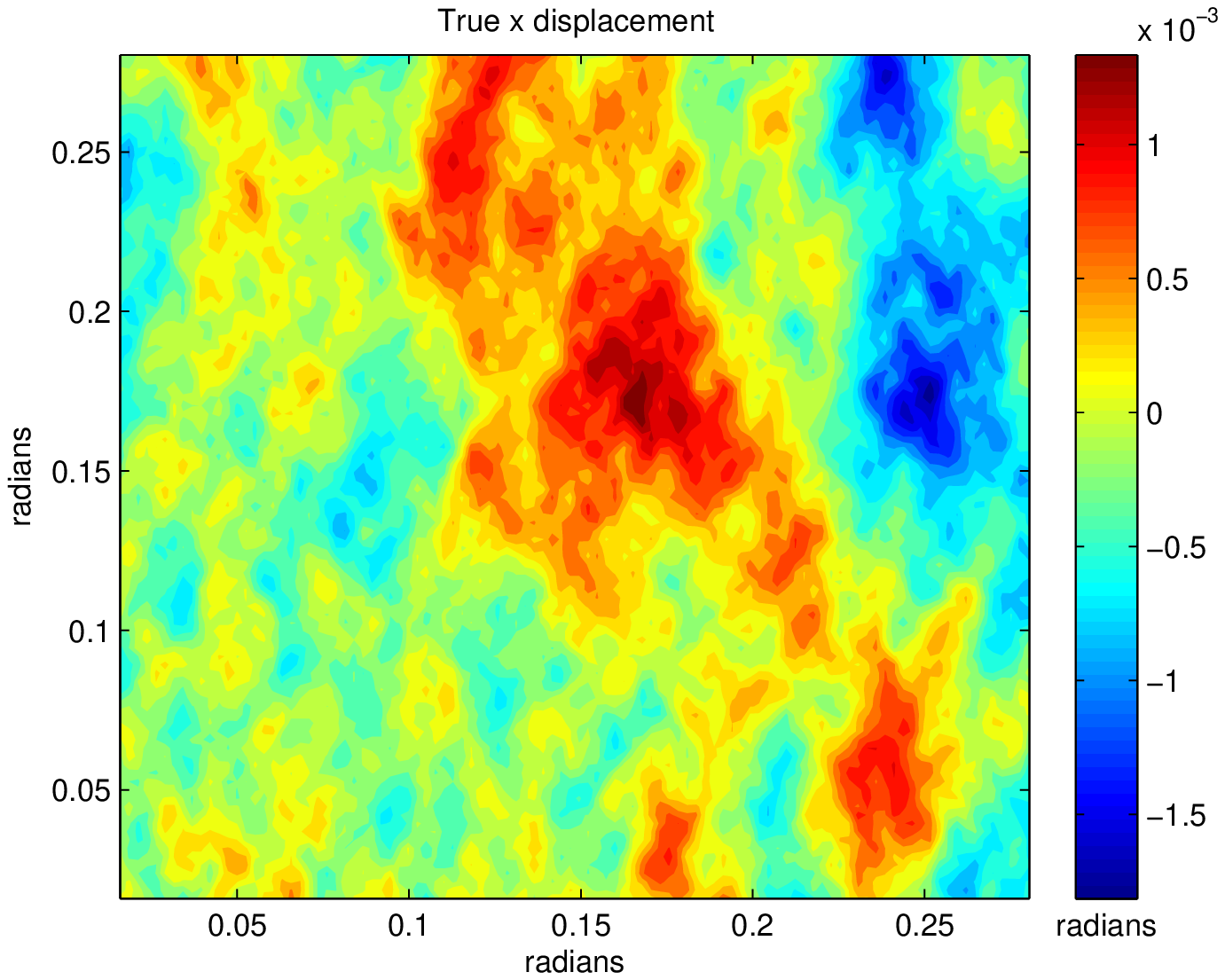}
\caption { The right diagram shows $\phi_x$,  where $\nabla\phi\equiv (\phi_x,\phi_y)$ is the true gravitational displacement field used in the simulation. The left diagram shows the estimate $\hat\phi_x$ which is derived from the local quadratic estimates using the methodology described in Section \ref{Rec}. \label{EstX} }
\end{figure*}

\subsection{Reconstructing $\phi$ from $\hat q^\phi$}
\label{Rec}

When observing the full sky, the estimates of $\kappa$ will allow one to recover the gravitational potential $\phi$ by solving the poisson equation $\nabla^2 \phi=-2\kappa$ (up to a constant). With partial sky observations, however, the shear is needed to break ambiguity corresponding to different boundary conditions.  We do this in two stages, first using $\hat q^{\phi}_{xx}$, $\hat q^{\phi}_{xy}$ and $\hat q^{\phi}_{yy}$ (regarded as functions of the local neighborhood midpoint $\bs x_0$) to recover the estimated displacement field $(\hat \phi_{x},\hat\phi_{y})$, then using this displacement field to recover the estimated potential $\hat\phi$. 
 To handle this, we adopt the method of  \cite{Simc1990,horn90} and define $\hat \phi_{x},\hat\phi_{y}$ as minimizers of functionals $F_1$ and $F_2$ defined as
\begin{align*}
F_1(\phi_x)&\equiv  \int dxdy \Bigl[(\phi_{xx}-\hat q^{\phi}_{xx})^2+ (\phi_{xy}-\hat q^{\phi}_{xy} )^2 \Bigr]  \\
F_2(\phi_y)&\equiv   \int dxdy \Bigl[(\phi_{xy}-\hat q^{\phi}_{xy} )^2+ (\phi_{yy}-\hat q^{\phi}_{yy} )^2 \Bigr].
\end{align*}
In particular, $\hat \phi_x$ satisfies $F_1(\hat\phi_x)=\min_{\phi_x} F_1(\phi_x)$ and similarly for $\hat\phi_y$.
See \cite{Agra06} for details of the minimization algorithm. 
 Now we use the estimated displacements $(\hat \phi_{x},\hat \phi_{y})$ to define the estimated potential $\hat\phi$ as the minimizer of the functional $F_3$ defined as
 \[F_3(\phi)\equiv   \int dxdy \Bigl[ (\phi_x-\hat \phi_{x} )^2+ (\phi_y-\hat \phi_{y} )^2\Bigr].\]
The minimization is needed to account for the fact that our estimates are noisy versions of the truth and therefore may not correspond to an integral vector field for which a potential exists. A consequence is that the estimate $\hat\phi$ is \lq shrunk' towards zero when the algorithm fits a gradient to a vector field which may have non vanishing curl. This shrinking can be seen in Fig.~\ref{fig2} looking at the scatter plot of green points. These points show  $(\nabla^2 \phi(\bs x_0), \nabla^2 \hat \phi(\bs x_0))$ for each local neighborhood $\mathcal N_\delta(\bs x_0)$. One can clearly see the shrinkage effect by noticing the slope of the trend in the green points is less than one.  We  undo this shrinkage effect by multiplying $\hat \phi$ by a factor that undoes this shrinkage. The multiplication factor, denoted $c$, is determined by matching the variance of the raw estimates $\nabla^2 \hat q^\phi$ with $c\nabla^2 \hat \phi$. The result of this correction factor is seen in the scatter plot of the red points, in Fig.~\ref{fig2}, which show the local convergence estimates versus truth after the correction factor $(\nabla^2 \phi,  c \nabla^2 \hat \phi-0.2)$.

\begin{figure*}
\includegraphics[height=6.7cm]{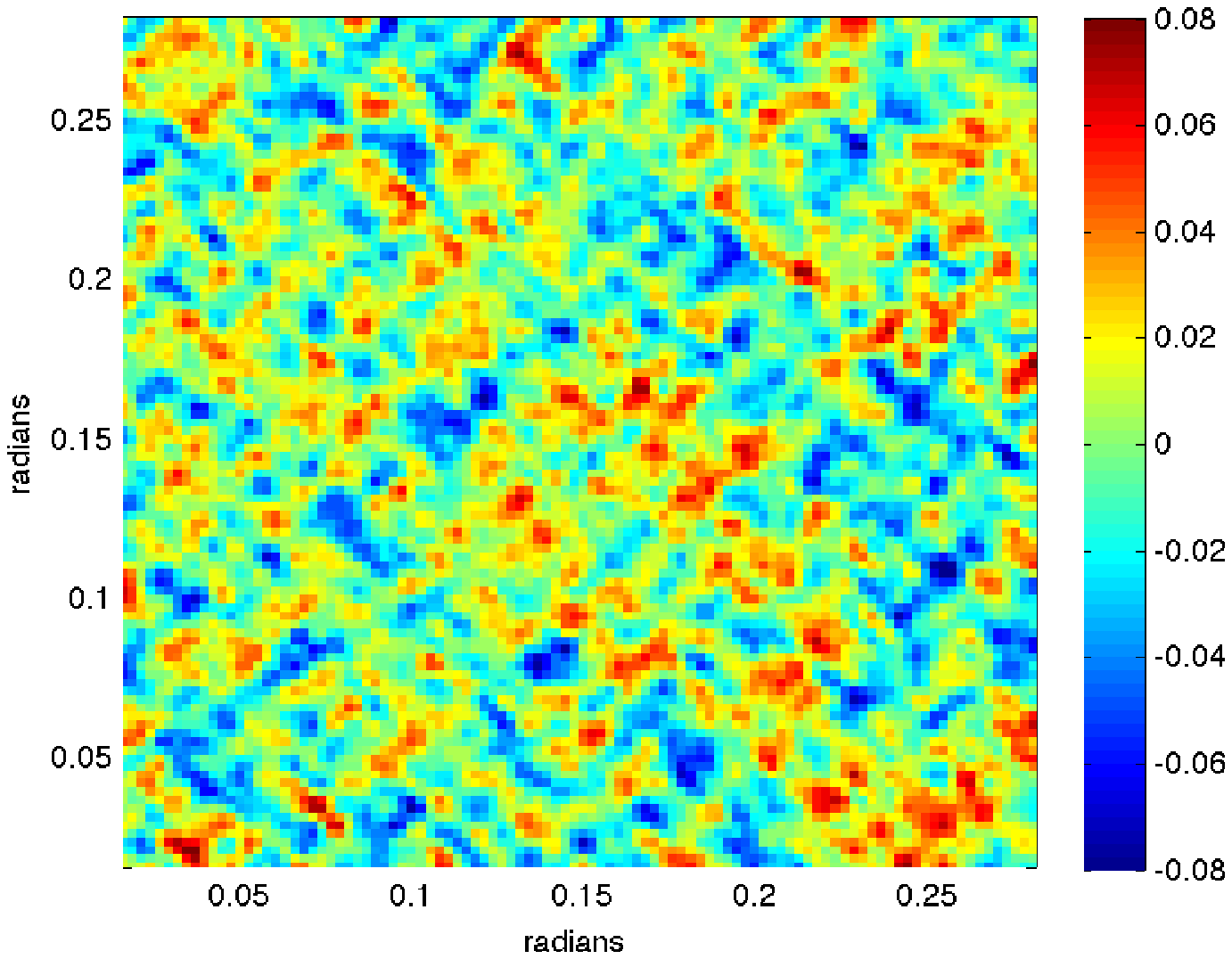}\\
\includegraphics[height=6.7cm]{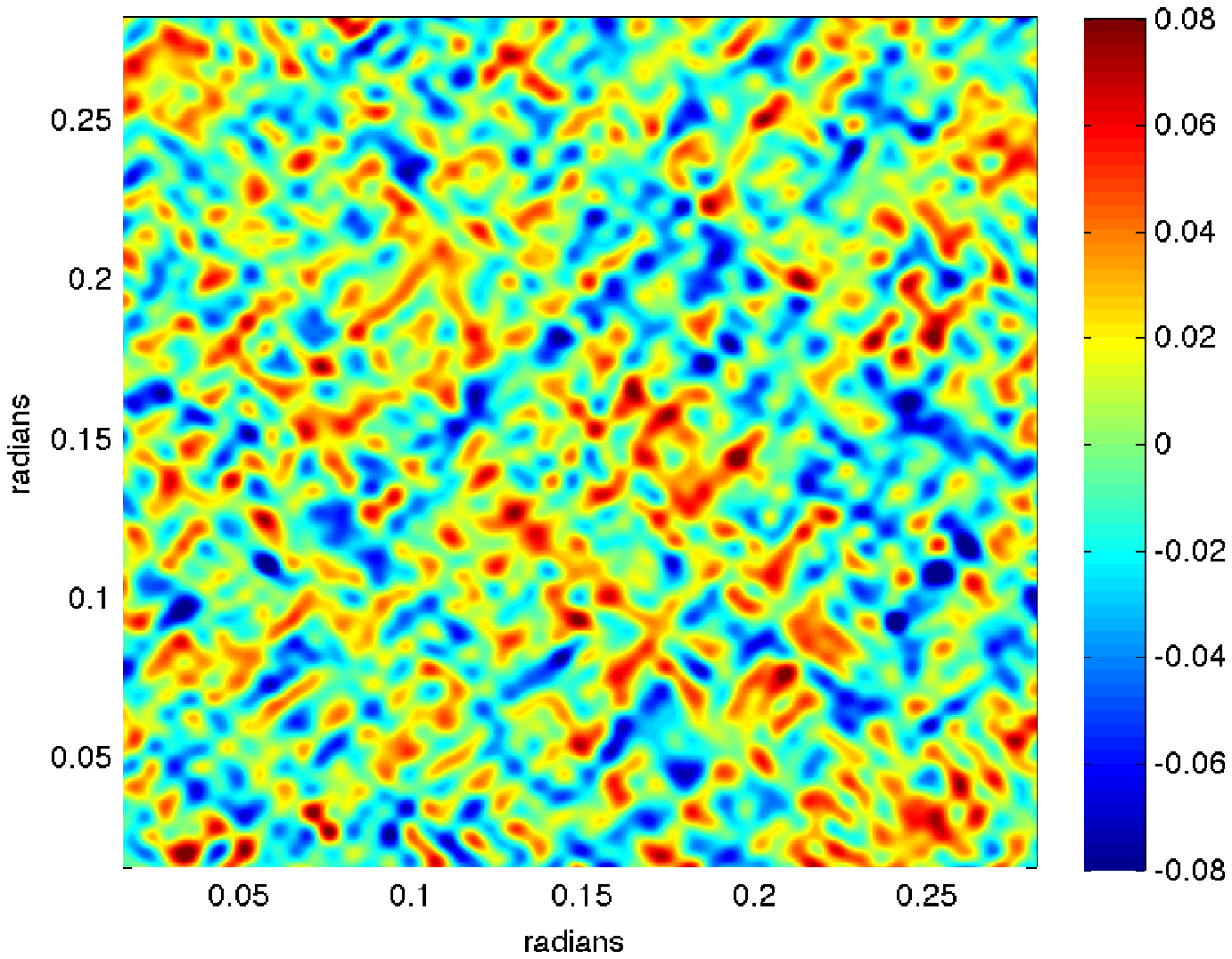} \includegraphics[height=6.7cm]{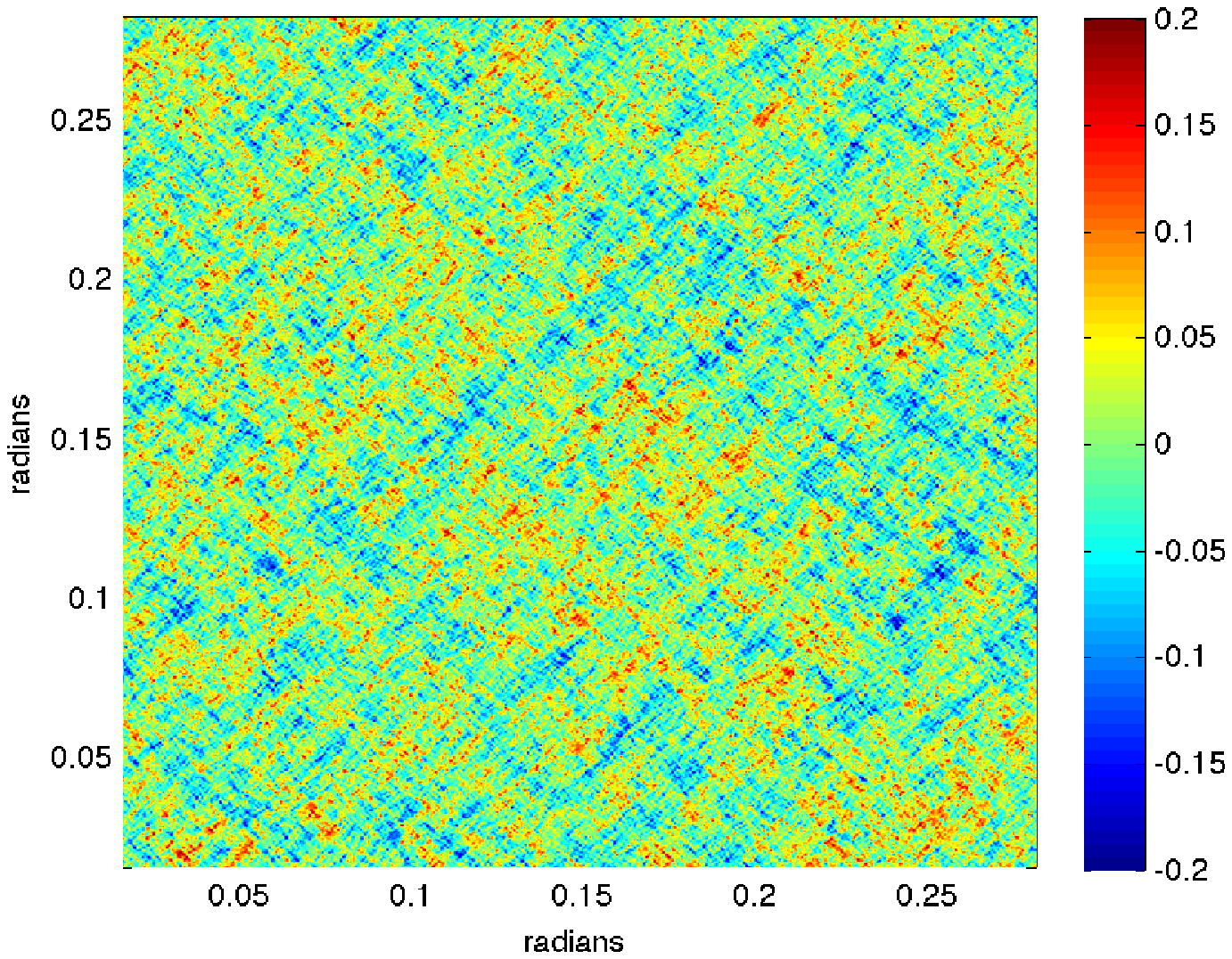}
\caption{ The top diagram shows the estimate of $\phi_{xy}$ (which corresponds to minus the imaginary part of the shear $\gamma$) where $\phi$ denotes the gravitational potential. The bottom two diagrams show the simulation truth: bottom left shows $\phi^\text{lp}_{xy}$ where $\phi^\text{lp}$ denotes the low pass filter  $\phi^\text{lp}(\bs \ell)\equiv   \min\Bigl\{1, \Bigl[2-\frac{0.006}{\pi} |\bs \ell|  \Bigr]^+\Bigr\} \phi(\bs \ell)$ (see Section \ref{pprior} for a discussion); bottom right shows $\phi_{xy}$. Notice that the estimate of $\phi_{xy}$ tracks the low pass filter $\phi^\text{lp}_{xy}$  and does not have the high frequency behavior seen in the simulation truth $\phi_{xy}$. \label{bbb}}
\end{figure*}

The estimated $\hat \phi_{x}$  (after correcting for the shrinkage)  along with the true displacement $\phi_x$ (used in the simulation) are shown in Fig.~\ref{EstX}. The estimated $\hat \phi$ along with the true gravitational potential $\phi$ are shown in Fig.~\ref{fig1}.  These two figures demonstrate accurate reconstruction of both the gravitational potential and the displacement field.  In addition, by differentiating the estimated potential, $\hat\phi$, one obtains smoothed estimates of convergence and shear (smoothed from the fitting of $\hat\phi$). In Fig.~\ref{bbb} we plot  the estimate  $\hat\phi_{xy}$ (which corresponds to minus the imaginary part of the shear $\gamma$), along with $\phi_{xy}$  (bottom right) and the low pass filter $\phi_{xy}^\text{lp}$ (bottom left) defined in Section \ref{pprior}. Notice that the estimate  $\hat\phi_{xy}$ tracks the derivatives of the low pass filter $\phi_{xy}^\text{lp}$, whereas the additional high frequency in $\phi_{xy}$ is not accurately estimated from  $\hat\phi_{xy}$. This is presumably due to the local fitting of a quadratic potential over the neighborhoods $\mathcal N_\delta (\bs x_0)$.

\section{Neighborhood size and structure}
\label{Infor}
We define the following measure of information which is used  as a metric for choosing the width of the neighborhood and other parameters of our estimation method:
\begin{align*} 
&\text{Information for $q^\phi$}\equiv \\
&\qquad\qquad\frac{\text{variance of the prior on $q^\phi$}}{\text{expected variance of the posterior on $q^\phi$}}. 
\end{align*}
The above information metric is essentially a measure of signal to noise ratio (squared). The variance of the prior corresponds to the squared magnitude of the signal, whereas the expected variance of the posterior is a proxy for the squared magnitude of the noise. We use simulations  to estimate this information  (while using the hessian of the posterior density at $\hat q^\phi$ to approximate posterior variance) and use it for guidance when choosing the tuning parameters for our estimation algorithm. Note: we avoided a lengthy and rigorous simulation study to choose global optimal tuning parameters, opting for a less rigorous simulation study which yields, potentially, sub-optimal but reasonable algorithmic parameters. 

The main  parameter that needs tuning is the local neighborhood size $\delta$.
Notice that our information measure attempts to balance two competing quantities when choosing a neighborhood size, the larger the neighborhood the smaller the signal $q^\phi$ (from the low pass filter). On the other hand, larger neighborhoods correspond to more data when the resolution is fixed.  Using this metric, $\delta=0.006$ radians (20.6 arcmin) emerges as a good neighborhood size when the beam FWHM
is $0.25$ arcmin and the noise levels are $\sqrt{2}$ and $1\, \mu K$-arcmin pixels for $Q,U$ and $T$, respectively.  

Due to computational limitations associated with larger neighborhoods we found it necessary to down-sample the local neighborhoods by discarding pixels.  Using the information metric we were able to isolate that randomly sampling the pixels  seemed preferable to evenly downsampling to a courser grid. Moreover, we found that using different randomly selected pixels for $T, Q$ and $U$ was preferable to using the same random pixels for all the Stokes fields. Therefore, for each local neighborhood we selected 300 random pixels in $\mathcal N_\delta(\bs x_0)$ for the $T$ observations, then randomly selected 300 pixels from those remaining  for $Q$ and finally 300 pixels from the remaining unselected pixels for $U$ (allowing overlaps when the local neighborhood size had fewer than $900$ pixels).

\section{Discussion}

We have demonstrated the feasibility of using a local Bayesian estimate  to accurately map the gravitational potential and displacement fields under low noise, small beam experimental conditions.  
  The motivation for developing this estimate stems, in part, from speculation that likelihood methods will allow superior mapping of the lensing structure (compared to the quadratic estimator) under low noise levels.  
   The main difference between the global estimates of  \cite{HiraSel2003a,HiraSel2003b} and the local estimate presented here is the nature of the likelihood approximation. In \cite{HiraSel2003a,HiraSel2003b} the global likelihood is defined as a functional on the unknown gravitational potential $\phi$ and approximations are made to this functional. Our method, in contrast, uses a nearly exact likelihood---exact up to approximation (\ref{CovFun}) in Appendix \ref{Newt}---but under a local modeling approximation that assumes a quadratic $\phi$. One advantage is the added precision available to model instrumental and foreground characteristics.  For example, the local analysis models the beam convolved CMB rather than the deconvolved CMB.  Deconvolution induces spatial correlation in the additive instrumental noise which is potentially nonstationary if the beam spatially varies. Since this noise is not invariant under warping it complicates the global likelihood.  Another advantage is that the local estimates are relatively easy to implement and parallelize.  In addition, the local estimate automatically uses the highest signal to noise combinations of $Q, U$ and $T$ so there is no need to re-derive the optimal quadratic combinations for different experimental conditions.

  The local analysis is not free from disadvantages however. A global analysis is presumably much better suited for estimating long wavelengths in the gravitational potential and wavelengths that are shorter than the local neighborhood size. Moreover, since our estimates are defined implicitly---as the maximum of the posterior density---it is difficult to derive expected error magnitudes.  However, the results presented here show that under some experimental conditions the advantages overcome the disadvantages.  
Moreover our local estimate uses an approximation that is inherently different from the Taylor approximation used to derive the quadratic estimator. This leaves open the possibility that the local estimate may have  different bias and error characteristics which could  compliment the quadratic estimator, rather than replace it.  

\appendix

\section{Simulation details}
\label{SimDets}

The fiducial cosmology used for the simulations is based on a flat,
power law $\Lambda$CDM cosmological model, with baryon density
$\Omega_b=0.044$; cold dark matter density $\Omega_\text{cdm}=0.21$;
cosmological constant density $\Omega_\Lambda=0.74$; Hubble parameter
$h=0.71$ in units of 100$\,$km$\,$s$^{-1}\,$Mpc$^{-1}$; primordial
scalar fluctuation amplitude $A_s(k=0.002\,$Mpc$^{-1}) = 2.45\times
10^{-9}$; scalar spectral index $n_s(k=0.002\,$Mpc$^{-1}) = 0.96$;
primordial helium abundance $Y_P=0.24$; and reionization optical depth
$\tau_r=0.088$. The CAMB code is used to generate the theoretical
power spectra \citep{CAMB}.

We start by simulating maps of the unlensed CMB Stokes parameters $T,Q,U$.  The following  Riemann sum approximation is used for the random fields $T,Q,U$
\begin{align}
\label{sim1}
T(\bs x)&\approx \sum_{\bs \ell}\frac{ Z^T_{\bs \ell}  \sqrt{\Delta{\ell_1}\Delta\ell_2} }{2\pi}   e^{i \bs x\cdot \bs \ell}  \sqrt{C^{T}_{\bs \ell}} \\
\label{sim2}
Q(\bs x)&\approx \sum_{\bs \ell}  \frac{Z^E_{\bs \ell}\sqrt{\Delta{\ell_1}\Delta\ell_2} }{2\pi}  e^{i \bs x\cdot \bs \ell} \cos(2\varphi_{\bs \ell})\sqrt{C^{E}_{\bs \ell}}   \\
\label{sim3}
U(\bs x)&\approx \sum_{\bs \ell}  \frac{Z^E_{\bs \ell}\sqrt{\Delta{\ell_1}\Delta\ell_2} }{2\pi}  e^{i \bs x\cdot \bs \ell} \sin(2\varphi_{\bs \ell})\sqrt{C^{E}_{\bs \ell}}  
\end{align}
where $\varphi_{\bs \ell}=\tan^{-1}(\ell_2/\ell_1)$; $\Delta\ell_1, \Delta\ell_2$ are the frequency spacing in the two coordinate directions;  for each $\bs \ell$, $Z_{\bs \ell}^T$ and $Z_{\bs \ell}^E$ are mean zero complex Gaussian random variables such that  $\langle Z_{\bs \ell}^T {Z^T_{\bs \ell^\prime}}^* \rangle=\langle Z_{\bs \ell}^E {Z^E_{\bs \ell^\prime}}^* \rangle= \delta_{\bs \ell-\bs \ell^\prime}$,  $\langle Z_{\bs \ell}^T {Z^E_{\bs \ell^\prime}}^* \rangle= \frac{C^{TE}_{\bs \ell}}{\sqrt{C^{T}_{\bs \ell}}\sqrt{C^{E}_{\bs \ell}}}\delta_{\bs \ell-\bs \ell^\prime}$,   $Z^T_{-\bs \ell}={Z^T_{\bs \ell}}^*$ and  $Z^E_{-\bs \ell}={Z^E_{\bs \ell}}^*$. To enforce the proper cross correlation between $Z^T_{\bs \ell}$ and $Z^E_{\bs \ell}$ we set 
\begin{equation}
 \Bigl[\begin{array}{c} Z^T_{\bs \ell}\\Z^E_{\bs \ell}\end{array}\Bigr]=\frac{1}{\sqrt{2}}
\left[ \begin{array}{cc} -\sqrt{1-\rho} & \sqrt{1+\rho}\\ \sqrt{1-\rho}&  \sqrt{1+\rho}  \end{array}\right]  \left[ \begin{array}{c}  W^1_{\bs \ell}\\ W^2_{\bs \ell}\end{array}\right]
\end{equation}
where $\rho\equiv \frac{C^{TE}_{\bs \ell}}{\sqrt{C^T_{\bs \ell}}\sqrt{C^E_{\bs \ell}}}$, and for each $\bs \ell$, $W_{\bs \ell}^1,W_{\bs \ell}^2$ are mean zero complex Gaussian random variables such that  $\langle W_{\bs \ell}^1 {W^1_{\bs \ell^\prime}}^* \rangle= \delta_{\bs \ell-\bs \ell^\prime}$,  $\langle W_{\bs \ell}^2 {W^2_{\bs \ell^\prime}}^* \rangle= \delta_{\bs \ell-\bs \ell^\prime}$, $\langle W_{\bs \ell}^1 {W^2_{\bs \ell^\prime}}^* \rangle=0$, $W^1_{-\bs \ell}={W^1_{\bs \ell}}^*$ and  $W^2_{-\bs \ell}={W^2_{\bs \ell}}^*$.

In our simulation, the above sums---equations (\ref{sim1}),(\ref{sim2}) and (\ref{sim3})---are taken over frequencies $\bs \ell\in \bigl\{ \frac{2\pi}{L}\bs k:\bs k\in \{-N/2,\ldots,N/2-1 \}^2 \bigr\}$ where $L=0.2967$ radians so that  $T$ will be  periodic on $[-L/2,L/2]^2$. The limit $N=L/\Delta_x$ is chosen to match the resolution in pixel space, denoted $\Delta_x$, so that FFT can be used to compute the sums  (\ref{sim1}),(\ref{sim2}) and (\ref{sim3}) which, after simplification, becomes
\begin{align}
T(\bs j\Delta_x)&\approx \sum_{\bs k}  Z^T_{\frac{2\pi}{L}\bs k} \frac{2\pi}{L}  e^{i 2\pi \bs k \cdot \bs j/N}\sqrt{C^{T}_{\frac{2\pi}{L}\bs k}} \\
Q(\bs j\Delta_x)&\approx \sum_{\bs k}  Z^E_{\frac{2\pi}{L}\bs k} \frac{2\pi}{L}  e^{i 2\pi \bs k \cdot \bs j/N}\cos(2\varphi_{\bs \ell})\sqrt{C^{E}_{\frac{2\pi}{L}\bs k}}\\
U(\bs j\Delta_x)&\approx \sum_{\bs k }  Z^E_{\frac{2\pi}{L}\bs k} \frac{2\pi}{L}  e^{i 2\pi \bs k \cdot \bs j/N}\sin(2\varphi_{\bs \ell})\sqrt{C^{E}_{\frac{2\pi}{L}\bs k}}
\end{align}
for each $\bs j\in \{ N/2,\ldots,N/2-1 \}^2$ where the sums range over $\bs k\in \{-N/2,\ldots,N/2-1 \}^2 $.
The matrix of values $\bigl[ T(\bs j\Delta_x)\bigr]_{\bs j\in \{-N/2,\ldots,N/2-1 \}^2}$, for example, can then be simulated by a two dimensional FFT of the matrix $\left[ Z^T_{\frac{2\pi}{L}\bs k} \frac{2\pi}{L}  e^{i 2\pi \bs k \cdot \bs j/N}\sqrt{C^{T}_{\frac{2\pi}{L}\bs k}} \right]_{\bs k\in \{-N/2,\ldots,N/2-1 \}^2}$. The identities $W^1_{-\bs \ell}={W^1_{\bs \ell}}^*$  and $W^2_{-\bs \ell}={W^2_{\bs \ell}}^*$ are enforced using a two dimensional FFT of two $N\times N$ matrices with independent standard Gaussian random entries.

{\em Remark:} 
Typically the above method suffers from an aliasing error when truncating to a finite sum in  (\ref{sim1}),(\ref{sim2}) and (\ref{sim3}). We avoid any such complication by setting the power spectrum in $C^T_{\bs \ell}, C^Q_{\bs \ell}$ and $C^U_{\bs \ell}$ to zero for all frequencies beyond $|\bs \ell|=6000$. We justify this truncation since both diffusion damping and the beam FWHM of $0.25^\prime$ combine to produce negligible amplitude in the CMB Stokes parameters at frequencies $|\bs \ell|\geq 6000$ compared  to the noise level.

{\em Remark:} Since the full sky Stokes parameters $T,Q,U$ are defined on the sphere, the theoretical power spectrum for $C^T_\ell$, $C^Q_\ell$, $C^U_\ell$ are only defined on integers $\ell$. Our flat sky approximation is obtained by extending $C^T_\ell$, $C^Q_\ell$ and $C^U_\ell$ to $\bs \ell\in\Bbb R^2$  by rounding the magnitude $|\bs \ell|$ to the nearest integer.
See Appendix C in \cite{Hu2000} for a derivation of this flat sky approximation.

To get a realization of the lensed CMB Stokes parameters $T,Q,U$ we use the above method to generate a high resolution simulation of $T,Q,U$ and the gravitational potential $\phi$ on a $17^o \times 17^o$ patch of the flat sky with $0.25$ arcmin pixels. The lensing operation is performed by taking the numerical gradient of $\phi$, then using linear interpolation to obtain the values $T(\bs x + \nabla \phi(\bs x)), Q(\bs x + \nabla \phi(\bs x)), U(\bs x + \nabla \phi(\bs x))$. The beam effect is then performed in Fourier space using FFT of the lensed fields. 
Finally, we down-sample the lensed fields, every $4^\text{th}$ pixel, to obtain the desired arcmin pixel resolution for the simulation output.

\section{Newton's method for maximizing the local posterior}
\label{Newt}
In this section we discuss our numerical procedure for maximizing the local posterior given by (\ref{FormPost}). We remark that  calculations need to be fast since they will be performed on each local neighborhood for which a shear and convergence estimate is required.
We discuss how the  FFT can be used to  to compute the covariance matrix, denoted $\Sigma_{q^\phi}+N$ in Section \ref{lllike}, and the corresponding derivatives with respect to the unknown coefficients of $q^\phi$. We let $M$ be the symmetric $2\times 2$ matrix defined as  $M\equiv \begin{pmatrix} 1+q_{xx}^\phi & q^\phi_{xy} \\ q^\phi_{xy} & 1+q^\phi_{yy}\end{pmatrix}$  so that  $\bs x+ \nabla q^\phi(\bs x)=M\bs x$. The matrix $M$ is regarded as the unknown which will be estimated from the data in the local neighborhood $\mathcal N_\delta(\bs x_0)$. Let $T_{b,M}$ denote a sheared temperature field, convolved with a Gaussian beam (with standard deviation $\sigma_b$) so that
\[T_{b,M}(\bs x)=\int_{\Bbb R^2} {d^2\bs y} T(M\bs x- M\bs y)  {e^{-{|\bs y|^2}/{(2\sigma^2_b)}}} ({\sigma_b^2\, 2 \pi })^{-1}  . \]
To compute the covariance matrix of the $T$ observations $\bs t=(t_1,\ldots,t_n)^\dagger$  in $\mathcal N_\delta(\bs x_0)$ (see equation (\ref{obbbsss})) one needs to evaluate  the following covariance function for a given test shear matrix $M$ at all vector lags $\bs h=\bs x_j-\bs x_k$ 
\begin{align*}
C_{T_{b,M}}( \bs h)&\equiv \langle T_{b,M}(\bs x+\bs h)T_{b,M}(\bs x) \rangle_\text{CMB}\\
&= \int_{\Bbb R^2} \frac{d^2\bs \ell}{(2\pi)^2} e^{i \bs \ell \cdot \bs h} e^{-\sigma_b^2 | \bs \ell|^2 } \frac{C^{T}_{M^{-1}\bs \ell}}{\det M}.
\end{align*}
All these calculations can be approximated using a FFT by noticing 
\begin{equation}
\label{CovFun}
C_{T_{b,M}}( \bs j\Delta_x) \approx \sum_{\bs k}  \frac{\Delta_\ell^2}{(2\pi)^{2}}  e^{i 2\pi \bs k \cdot \bs j/N}  e^{-\sigma_b^2 |\Delta_\ell \bs k|^2} \frac{C^{T}_{ \Delta_\ell M^{-1} \bs k}}{\det M} 
\end{equation}
where the sum ranges over $\bs k\in \{-N/2,\ldots,N/2-1 \}^2$, $\Delta_x$ is the pixel spacing, $\bs j \in \{-N/2,\ldots,N/2-1 \}^2$,  $\Delta_\ell=2\pi/L$ and $L=N\Delta_x$.  Then to compute the covariance between $t_j$ and $t_k$ we simply select the entry of the matrix $\left[ C_{T_{b,M}}( \bs j\Delta_x) + \sigma^2_T \delta_{j_1j_2}\right]_{\bs j\in \{-N/2,\ldots,N/2-1 \}^2}$ such that $\bs j\Delta_x=\bs x_j-\bs x_k$ (which was obtained by a single FFT).
  A similar technique can be used to compute all other covariance and cross-covariances among $T, Q$ and $U$ to construct the covariance matrix $\Sigma_{q^\phi}$.
 We remark that to speed up the computations we choose a smaller $N$ then the one used in the simulations ($N=4096$ in the simulation but  $N=256$ for the approximation of $C_{T_{b,M}}( \bs j\Delta_x)$).

Once the covariance matrix $\Sigma_{q^\phi}+N$ is constructed using the approximation (\ref{CovFun}) (and the analogous approximations for $Q,U$ and all cross correlations) the posterior is easily computed as $p(q^\phi | \bs z)\propto e^{\mathcal L(q^\phi|\bs z)}\pi(q^\phi)$ where $\mathcal L$ denotes the log likelihood (\ref{MMTT}) and $\pi$ is the prior distribution derived in Section \ref{pprior}. In principle, one can now simply use pre-existing minimization algorithms for maximizing the posterior  $p(q^\phi | \bs z)$ with respect to $q^\phi$. If one desires a more sophisticated Newton type algorithm for maximizing the posterior one often needs to compute the gradient and hessian of the posterior. Using the techniques of {\it automatic differentiation} (see \cite{neid}, for example) one can easily compute such derivatives if one can compute the rates of change of the covariance $\Sigma_{q^\phi}$ with respect the the elements of $M$. 

We finish this Appendix by noticing that the FFT can be used to approximate the derivatives of $\Sigma_{q^\phi}$ with respect to the elements of the matrix $M$, denoted $M_{k,j}$ for $k,j\in\{1,2\}$.  For illustration we focus on the covariance structure of the temperature field $T$ and mention that the extension to $Q,U$ is similar. First notice that by transforming variables $\bs \ell^\prime=M^{-1}\bs \ell$ one gets
\begin{align*}
\frac{dC_{T_{b,M}}( \bs h)}{dM_{k,j}}  &= \frac{d}{dM_{k,j}}\int \frac{d^2\bs \ell^\prime}{(2\pi)^2}e^{i (M\bs \ell^\prime) \cdot \bs h- \sigma_b^2|M\bs \ell^\prime|^2} {C^T_{\bs \ell^\prime}} \\
&=\int \frac{d^2\bs \ell^\prime }{(2\pi)^2}   e^{i (M\bs \ell^\prime) \cdot \bs h- \sigma_b^2|M\bs \ell^\prime|^2} C^T_{\bs \ell^\prime}\\
&\qquad\qquad\times \frac{d}{d M_{k,j}}   \Bigl[i (M\bs \ell^\prime) \cdot \bs h- \sigma_b^2|M\bs \ell^\prime|^2 \Bigr]\nonumber
\end{align*}
Now $d \Bigl[i (M\bs \ell^\prime) \cdot \bs h- \sigma_b^2|M\bs \ell^\prime|^2 \Bigr]\Bigl/ {d M_{k,j}} $ can be written as a sum $\sum_{k} c_k(\bs h) g_k(M, \bs \ell^\prime)$ so that  by re-transforming variables to $\bs \ell=M\bs \ell^\prime$ one gets
\begin{align*} 
\frac{dC_{T_{b,M}}( \bs h)}{dM_{k,j}}  &=\sum_k c_k(\bs h) \int \frac{d^2\bs \ell  }{(2\pi)^2} e^{i \bs \ell \cdot \bs h- \sigma_b^2|\bs \ell|^2} \\
&\qquad\qquad\times{g_k(M, M^{-1}\bs \ell) } \frac{C^T_{M^{-1}\bs \ell}}{\det M}.  
\end{align*}
The point is that the above integrals can now be approximated using FFT to approximate the matrix of values $\Bigl[  \frac{dC_{T_{b,M}}( \bs j\Delta_x)}{dM_{k,j}}\Bigr]_{\bs j\in \{-N/2,\ldots,N/2-1 \}^2}$. The same method applies to approximate all higher order derivatives of the covariance matrix. 
These derivatives can then be used in a Newton type algorithm for finding the maximum {\it a-posteriori} estimates $\hat q^\phi$.

\nocite{*}


\begin{thebibliography}{10}


\bibitem{Agra06} Agrawal, A., Raskar, R. \& Chellappa, R. 2006, Lecture Notes in Computer Science, Springer Berlin/Heidelberg, 3951, 578-591

\bibitem{buncher} Bucher, M., Carvalho, C. S., Moodley, K., Remazeilles, M., arXiv:1004.3285 (2010)

\bibitem{carv}  Carvalho, C. S., Moodley, K., Phys. Rev. D 81, 123010 (2010)

\bibitem{dod} Dodelson, S., {\em Modern cosmology}, Academic Press (2003)

\bibitem{Guzik2000} Guzik, J., Seljak, U. \& Zaldarriaga, M., Phys. Rev. D 62, 043517 (2000)

\bibitem{Han2010} Hanson, D., Challinor, A., Efstathiou, G., Bielewicz, P., arXiv:1008.4403 (2010)

\bibitem{Hira2008} Hirata, C.~M., {Ho}, S., {Padmanabhan}, N., {Seljak}, U. \& 
	{Bahcall}, N.~A.. Phys. Rev. D 78, 043520 (2008)

\bibitem{HiraSel2003a} Hirata, C., \& Seljak, U., Phys. Rev. D 67, 043001 (2003a)


\bibitem{HiraSel2003b} Hirata, C., \& Seljak, U., Phys. Rev. D 68, 083002 (2003b)
%

\bibitem{horn90} Horn, B. 1990, Int'l J. Computer Vision, 5, 37-75

\bibitem{Hu2000} Hu, W., Phys. Rev. D 62, 043007 (2000)


\bibitem{Hu2001b} Hu, W., ApJ 557: L79-L83  (2001)




\bibitem{HuOka2002} Hu, W., \& Okamoto, T., ApJ 574: 566-574 (2002)

\bibitem{Kami} Kamionkowski, M., Kosowsky, A., Stebbins, A., Phys. Rev. D 55, 7368-7388 (1997)

\bibitem{Kaplin} Kaplinghat, M., Knox, L., Song, Y., Phys. Rev. Lett. 91, 241301 (2003)

\bibitem{Kesden} Kesden, M., Cooray, A., Kamionkowski, M., Phys. Rev. Lett. 89, 011304 (2002)

\bibitem{knox2002} Knox, L., Song, Y., Phys. Rev. Lett. 89, 011303 (2002)

\bibitem{CAMB}  {{Lewis}, A. and {Challinor}, A. and {Lasenby}, A.}, ApJ, 538: 473-476 (2000)

\bibitem{neid} Neidinger, R., SIAM Review, Vol. 52, No. 3, pp.545-563 (2010)

\bibitem{OkaHu2003}Okamoto, T., \& Hu, W., Phys. Rev. D 67, 083002 (2003)

\bibitem{SelH} Seljak, U, \& Hirata, C., Phys. Rev. D 69, 043005 (2004)

\bibitem{SelZlad1998}  Seljak, U., \& Zaldarriaga, M., arXiv:astro-ph/9805010 (1998)

\bibitem{SelZlad1999}  Seljak, U., \& Zaldarriaga, M., Phys. Rev. Letters, 82, 13, 2636-2639 (1999)



\bibitem{Simc1990} Simchony, T., Chellappa, R.,  \&
 Shao, M.   1990, IEEE Trans. Pattern An. Machine Intell. , 12, 435-446


\bibitem{Smth2006} Smith, K., Hu W., Manoj, K., Phys. Rev. D 74, 123002 (2006)

\bibitem{Smith2007} Smith, K., Zahn, O. \& Dore, O., Phys. Rev. D 76, 043510 (2007)



\bibitem{ZaldSel1999} Zaldarriaga, M., \& Seljak, U., Phys. Rev. D 59, 123507 (1999)


\bibitem{Zald2000} Zaldarriaga, M., Phys. Rev. D, 62 (2000)


\end{thebibliography}

\end{document}